\documentclass[a4paper,11pt]{article}
\usepackage{aaskaiid}
\usepackage{scalerel}
\usepackage{xcolor}
\usepackage{ulem} 
\usepackage{enumitem}
\usepackage{array}
\usepackage{orcidlink}

\newcommand\HII{H{\sc ii}}
\newcommand\HI{H{\sc i}}\newcommand{\blos}{$\bf{B_{\rm{LOS}}}$}
\newcommand{\bpos}{$B_{\rm{POS}}$}

\newcommand*\arcsec{\ensuremath{^{\prime\prime}}}

\title{3D Magnetic Field Vectors in Space: Bubbles, Clouds, and Filaments}
\ShortTitle{3D Fields}

\author[1,2]{Mehrnoosh Tahani\orcidlink{0000-0001-8749-1436}}
\ShortName{Tahani et al.} 
\author[3,4,5]{Anna Ordog\orcidlink{0000-0002-2465-8937}}
\author[3, 6]{Jennifer West\orcidlink{0000-0001-7722-8458}}
\author[7]{Georgia V. Panopoulou\orcidlink{0000-0001-7482-5759}}
\author[8,9]{Hiroko Shinnaga\orcidlink{0000-0001-9407-6775}}
\author[10]{Marijke Haverkorn\orcidlink{0000-0002-5288-312X}}

\affiliation[1]{Department of Physics \& Astronomy, University of South Carolina, Columbia, SC 29208, USA}
\affiliation[2]{Kavli Institute for Particle Astrophysics \& Cosmology (KIPAC), Stanford University, Stanford, CA 94305, USA}
\emailAdd{mtahani@mailbox.sc.edu}
\affiliation[3]{Dominion Radio Astrophysical Observatory, Herzberg Research Centre for Astronomy and Astrophysics, National Research Council Canada, PO Box 248, Penticton, BC, V2A 6J9, Canada}
\emailAdd{aordog@uwo.ca}
\affiliation[4]{Department of Computer Science, Math, Physics, \& Statistics, University of British Columbia, Okanagan Campus, Kelowna, BC V1V 1V7, Canada}
\affiliation[5]{Department of Physics \& Astronomy, University of Western Ontario, 1151 Richmond Street, London, ON, N6A 3K7, Canada}
\affiliation[6]{
School of Natural Sciences, University of Tasmania, PO Box 807, Sandy Bay, TAS 7006, Australia}
\emailAdd{jennifer.west@nrc-cnrc.gc.ca}
\affiliation[7]{Department of Space, Earth and Environment, Chalmers University of Technology, 412 93, G\"{o}teborg, Sweden}
\emailAdd{georgia.panopoulou@chalmers.se}
\affiliation[8]{Department of Physics and Astronomy, Graduate School of Science and Engineering, Kagoshima University, 1-21-35 Korimoto, Kagoshima, Kagoshima 890-0065, Japan}
\affiliation[9]{Amanogawa (Milkyway) Galaxy Astronomy Research Center (AGARC), Graduate School of Science and Engineering, Kagoshima University, 1-21-35 Korimoto, Kagoshima, Kagoshima 890-0065, Japan}
\emailAdd{shinnaga@sci.kagoshima-u.ac.jp}
\affiliation[10]{Department of Astrophysics/IMAPP, Radboud University, Nijmegen, PO Box 9010, 6500 GL Nijmegen, the Netherlands}
\emailAdd{m.haverkorn@astro.ru.nl}

\abstract{Magnetic fields play important roles in the star-formation process across different spatial scales. The interplay between magnetic field strength (a key component of the interstellar medium’s energy budget) and field orientation relative to density structures impacts how interstellar material evolves toward star formation. To understand galactic evolution toward stars, planets, and ultimately life, we need to map three-dimensional (3D) magnetic field vectors in 3D space. However, determining full vector information remains challenging due to projection effects and the complex relationship between observable tracers and field geometry. We outline the observational techniques that can be used to probe the 3D magnetic field structures of objects such as supernova remnants (SNR), superbubbles, \HII\ regions, and \HI\ filaments in the diffuse interstellar medium (ISM), and objects in the dense ISM such as molecular clouds, filaments, and cores. The main SKA-specific observational techniques include synchrotron emission and Faraday rotation of both compact sources and the diffuse emission. We discuss how SKA AA4 will allow implementation of the techniques we describe, leveraging the vastly improved sensitivity, resolution and uv-coverage compared to existing datasets. This will enhance our ability to reconstruct 3D magnetic field vectors, advancing our understanding of magnetic fields in Galactic evolution and star formation.
}

\begin{document}
\maketitle
\newcommand{\actaa}{Acta Astron.} 
\newcommand{\araa}{ARA\&A} 
\newcommand{\aar}{A\&ARv} 
\newcommand{\aapr}{A\&ARv} 
\newcommand{\ab}{Astrobiol.} 
\newcommand{\aj}{AJ} 
\newcommand{\apj}{ApJ} 
\newcommand{\apjl}{ApJL} 
\newcommand{\apjs}{ApJSS} 
\newcommand{\ao}{Appl. Opt.} 
\newcommand{\apss}{Astro. \& Space Sci.} 
\newcommand{\aap}{A\&A} 
\newcommand{\aaps}{A\&AS.} 
\newcommand{\baas}{Bull. Am. Astron. Soc.} 
\newcommand{\caa}{Chinese A\&A} 
\newcommand{\cjaa}{Chinese J. A\&A} 
\newcommand{\cqg}{Class. Quantum Gravity} 
\newcommand{\gal}{Galaxies} 
\newcommand{\gca}{Geo. Cosmo. Acta} 
\newcommand{\icarus}{Icarus} 
\newcommand{\jcap}{JCAP} 
\newcommand{\jgr}{J. Geophys. Res.} 
\newcommand{\jgrp}{J. Geophys. Res. Planets} 
\newcommand{\jqsrt}{JQSRT} 
\newcommand{\memsai}{Mem. SAIt} 
\newcommand{\mnras}{MNRAS} 
\newcommand{\nat}{Nature} 
\newcommand{\nastro}{Nat. Astron.} 
\newcommand{\ncomms}{Nat. Commun.} 
\newcommand{\nphys}{Nat. Phys.} 
\newcommand{\na}{New Astron.} 
\newcommand{\nar}{New Astron. Rev.} 
\newcommand{\physrep}{Phys. Rep.} 
\newcommand{\pra}{Phys. Rev. A} 
\newcommand{\prb}{Phys. Rev. B} 
\newcommand{\prc}{Phys. Rev. C} 
\newcommand{\prd}{Phys. Rev. D} 
\newcommand{\pre}{Phys. Rev. E} 
\newcommand{\prl}{Phys. Rev. L.} 
\newcommand{\psj}{Planet. Sci. J.} 
\newcommand{\planss}{Planet. Space Sci.} 
\newcommand{\pnas}{Proc. Natl Acad. Sci. USA} 
\newcommand{\procspie}{Proc. SPIE} 
\newcommand{\pasa}{PASA} 
\newcommand{\pasj}{PASJ} 
\newcommand{\pasp}{PASP} 
\newcommand{\rmxaa}{RMXAA} 
\newcommand{\sci}{Science} 
\newcommand{\sciadv}{Sci. Adv.} 
\newcommand{\solphys}{Sol. Phys.} 
\newcommand{\sovast}{Soviet Ast.} 
\newcommand{\ssr}{Space Sci. Rev.} 
\newcommand{\uni}{Universe} 
\newcommand{\an}{Astronomical Notes} 

\section{Introduction}

Polarized light surrounds us and can be utilized to reveal hidden information about our world, from the 3D glasses in movie theaters to the navigation systems of dragonflies and the hunting strategies of mantis shrimp \citep{Dalyetal2016MantisShrimp}. This same property of light serves as a powerful tool for mapping the invisible magnetic fields that help shape galaxies, stars, and planets throughout the universe. 

These magnetic fields compete with gravity and turbulence to control how matter flows through space, influencing when and where stars are born \citep{Pattleetal2023PP7}. The complexity of these interactions has made star formation one of the most pivotal unanswered questions in astrophysics. Understanding how the first stars formed, how galaxies evolve, and how planets emerge requires comprehending how stars form in today's Universe. Our galaxy provides an excellent laboratory for this fundamental investigation.

What we know about magnetic fields reveals their dual nature: they play complex and sometimes contradictory roles in star formation \citep{HennebelleInutsuka2019}. At times they impede the process, reducing star formation rates and efficiencies by providing magnetic pressure support against gravitational collapse \citep{McKee2007ARA&A..45..565M}.  However, in the final stages of stellar birth, magnetic fields become essential, enabling star formation through magnetic braking that removes excess angular momentum from rapidly spinning protostellar cores \citep{WursterLi2018FrASS...5...39W}. To fully understand these competing roles, we must map three-dimensional (3D) magnetic field vectors in 3D space and examine their relationship to cosmic and interstellar structures. 

\subsection{The necessity of 3D magnetic field vector measurements and the challenges in obtaining them}

\begin{figure}
\centering
\includegraphics[width=0.5\textwidth]{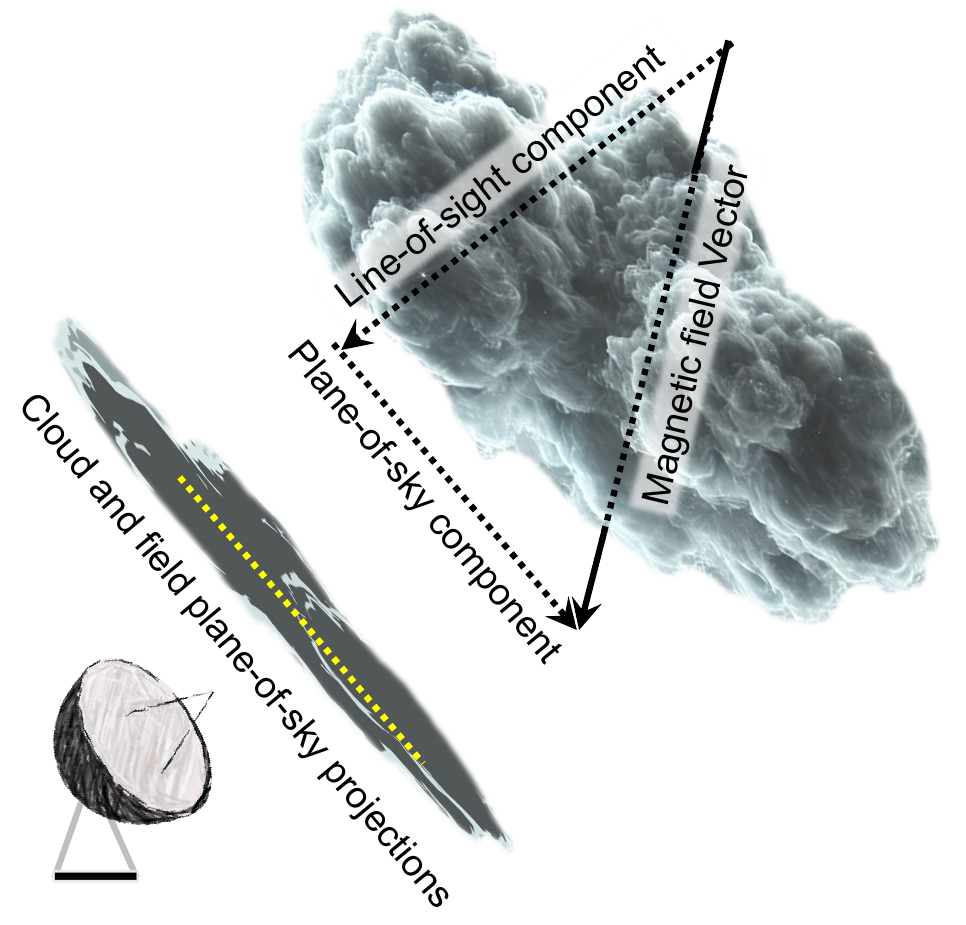}
\caption{Schematic illustration of the limitation of 2D projections. a perpendicular magnetic field vector to an elongated cloud may appear parallel when projected onto the plane of sky.}
\label{fig:2DLimitationCartoon}
\end{figure}

The 3D geometry of magnetic fields relative to 3D density structures can determine how these structures evolve \citep[e.g.][]{Tritsis2015MNRAS.451.4384T,Tahanietal2022O}. For example, consider a filamentary cloud: magnetic fields oriented parallel to the filament's long axis will preferentially resist radial gravitational collapse, while fields oriented perpendicular to the axis will resist collapse along the filament (poloidal collapse). These different field geometries can lead to distinct fragmentation modes and, ultimately, different populations of forming stars and star-formation rates \citep{SeifriedWalch2015}. Some studies~\citep{FiegePudritzI2000, FiegePudritzII2000, Inoueetal2018} suggest  that magnetic fields, depending on their morphology can allow for formation of more massive molecular clouds and stabilize the cloud against gravity and fragmentation.  Such critical distinctions are lost when we observe only two-dimensional projections \citep{Seifriedetal2020}, where nearly perpendicular structures can appear parallel when projected onto the plane of the sky as shown in Figure~\ref{fig:2DLimitationCartoon} \citep{Tahani2022}.

In addition, full 3D magnetic field vectors are crucial for studies of energy budgets in clouds. Magnetic energy density depends on the total magnetic field strength, and therefore only 3D magnetic field measurements can provide magnetic energy density values for comparison with thermal and turbulent energy densities.  Most observational measures of magnetic fields only provide a lower limit on the total field strength \citep[e.g.][]{Crutcher2012}, and often rely on assumptions to estimate the gas density or to infer the turbulent kinetic energy (such as isotropic turbulence).

Mapping 3D magnetic field vectors from observations is extremely challenging. The biggest challenge is obtaining the complete vector direction of the field, including the plane-of-sky direction. Other challenges include determining the 3D field strength and localizing the measurement in 3D space, since often probes of the plane-of-sky (POS) or line-of-sight (LOS) component are obtained from line-of-sight integrated observations, which may trace different phases of the ISM. 
To date, only two studies have successfully reconstructed full 3D magnetic field vectors in the interstellar medium \citep{Tahanietal2022O, Tahanietal2022P}, demonstrating both the difficulty of this endeavor and its importance.  These reconstructions revealed that 3D magnetic field vectors can trace the formation history and evolutionary path of molecular clouds and identify objects hidden to total emission observations (also see \citet{Mohammedetal2024} where Faraday tomography reveals an unidentified structure).

The most promising way to achieve 3D field vectors requires the incorporation of observations and techniques across the entire electromagnetic spectrum. In the following subsections, we describe the commonly-used observational techniques for probing interstellar magnetic fields and discuss how their integration enables 3D magnetic field reconstruction.

\subsection{SKA-Specific Common Observing Techniques}
\label{sec:ska-methods}

The SKA will enable measurements of interstellar magnetic fields with several different methods. The volume, coverage, and sensitivity of the expected data will be transformational in terms of our ability to reconstruct the magnetized ISM. The three main observables of magnetic fields from SKA polarization data are (a) Galactic synchrotron emission, (b) Faraday rotation of extended emission and background sources and (c) circular polarization of spectral lines due to the Zeeman effect. We briefly introduce each below.

\vspace{-3pt}
\subsubsection{Synchrotron Emission}

Synchrotron emission occurs when relativistic cosmic ray electrons undergo gyromotion around magnetic field lines. Synchrotron emission is commonly observed in regions of shocks, supernova remnants, and star-forming regions, and can be used to probe different magnetic field components. The emission is strongly polarized perpendicular to the projected plane-of-sky magnetic field component (\bpos ), as the radiation is emitted in the direction of the electron's acceleration. Therefore, the orientation of observed linear polarization reveals the \bpos\  orientation (with 180$^{\circ}$ ambiguity) in synchrotron-emitting regions integrated along the line of sight (in the absence of Faraday rotation). 
At higher frequencies ($>5$~GHz), Faraday rotation effects can be negligible depending on the environment and field strength, while at lower frequencies, observed Faraday rotation (see following section) can be used to de-rotate the polarization angles.

Synchrotron intensity can provide estimates of magnetic field strength under the assumption of equipartition between magnetic and cosmic ray energy densities. However, this assumption lacks direct observational validation and has been challenged by numerical simulations~\citep{Dacunhaetal2025CREquipartition}, which find that this assumption can lead to significant overestimation of magnetic field strength. \citet{Linzeretal2025EquipartitionCR} found that equipartition can be assumed at kpc scales but breaks down at smaller scales. Under the equipartition assumption, synchrotron total intensity may provide estimates of the overall field strength, while the ratio of polarized to total intensity gives information on the ratio of turbulent to uniform magnetic field components~\citep{BeckK2005, Borlaffetal2023}.

\subsubsection{Faraday rotation and Rotation Measure Grids}
\label{sec:RM}

Faraday rotation causes the polarization angle of linearly polarized emission passing through a magnetized region with free electrons to rotate. The amount of rotation can be quantified by the Faraday depth, $\phi$, as

\begin{equation}
	\frac{\phi (r)}{[\text{rad m}^{-2}]}=0.81\int_{\text{r}}^{\text{obs}} \frac{n_{\text{e}}}{[\text{cm}^{-3}]}\frac{B_{\rm{LOS}}}{[\mu G]}\frac{dl}{[\text{pc}]},
    \label{eq:RM}
\end{equation}
where $n_e$ is the thermal electron density, $B_{\rm{LOS}}$ is the strength of the magnetic field component along the line of sight, and $dl$ is the path length from the source at distance $r$ to the observer. When a background source of linearly polarized light (e.g. a pulsar or a quasar) propagates through a Faraday rotating foreground, the angle rotation $\Delta \theta$ is quadratic with observing wavelength and Faraday depth is referred to as Rotation Measure, RM~$= \Delta\theta / \Delta \lambda^2$.

When linearly polarized emission and Faraday rotation are intermixed, such as diffuse synchrotron emission and Faraday rotation in the interstellar medium, the Rotation Measure is undefined. In this case, multiple emission components with varying Faraday rotation contribute to the observed emission, which is now a complex function of observing wavelength. This can be described as a Faraday spectrum $F(\phi)$ in a technique called Faraday Tomography. Here, the complex observed polarization $P(\lambda^2) = Q(\lambda^2) + i U(\lambda^2)$ as a function of wavelength squared is Fourier transformed to \citep{Burn1966,BrentjensDeBruyn2005}:
\begin{equation}
F(\phi) \ = \int_{-\infty}^{+\infty}\ P(\lambda^2) \ 
e^{-2i \phi \lambda^2} \ d(\lambda^2).
\end{equation}
As wavelength coverage is limited and strictly positive in reality, the observed Faraday spectrum $\tilde{F}(\phi)$ is convolved with a Rotation Measure Spread Function (RMSF) $R(\phi)$, which is an artifact due to limited wavelength coverage\footnote{This is entirely analogous to the point spread function in radio interferometric data, which is an artifact due to limited spatial coverage of antennas.}. The observed Faraday spectrum is:
\begin{equation}
\tilde{F}(\phi) =  \frac{\int_{-\infty}^{+\infty} W(\lambda^2)P(\lambda^2) e^{-2i \phi \lambda^2} \ d\lambda^2 }{\int W(\lambda^2)d\lambda^2} = F(\phi) \star R(\phi),
\end{equation}
where $W(\lambda^2)$ is a window function denoting the wavelength coverage, $\star$ denotes convolution, and 
\begin{equation}
    R(\phi) = \int_{-\infty}^{+\infty}\ W(\lambda^2) \ 
e^{-2i \phi \lambda^2} \ d(\lambda^2).
\end{equation}
With intermixed synchrotron emission and Faraday rotation, the Faraday depth spectra can take on a range of different appearances, depending on both the physics in the ISM and the properties of the instrument used for the observations. A uniform magnetic field embedded in a uniform distribution of thermal and relativistic electrons will produce a slab-like feature in Faraday depth space, in which the largest Faraday depth corresponds to the furthest physical distance along the LOS. The scenario of multiple, discrete, synchrotron emitting and Faraday rotating regions or objects along the LOS will produce multiple features with differing intensities and polarization angles in the Faraday depth spectrum. The distances corresponding to the emission and Faraday rotation of these features can often be estimated by correlating the plane-of-sky structures in the polarized intensity at each Faraday depth with other tracers such as dust and \HI .

The observational signatures of extended synchrotron emission and Faraday rotation are heavily influenced by the observing parameters through the mechanism of depolarization. Depth depolarization arises from differing Faraday depths along the LOS yielding polarization angles that partially cancel out when they arrive at the telescope. Lower frequency observations are more susceptible to depth depolarization because Faraday rotation is more pronounced at longer wavelengths. Beam depolarization is caused by plane-of-sky variations in polarization angle (either intrinsic or a result of small-scale Faraday rotation variability) cancelling out within the beam of the telescope. Telescopes with larger beams are more prone to beam depolarization. Finally, bandwidth depolarization is the result of differential Faraday rotation within the frequency range encompassed by an observing channel of the instrument. The same frequency binning will cause more bandwidth depolarization at lower frequencies because of the wavelength-squared dependence of Faraday rotation. In general, depolarization limits the physical depths that are probed along the LOS by a given instrument, a phenomenon that has been referred to as the `polarization horizon' \citep{Uyaniker:2003, Hill:2018}. Differing amounts and types of depolarization between different instruments can be used to probe the same regions out to different depths.

\subsection*{Techniques using Faraday rotation}
Different techniques can help determine the magnetic fields associated with different environments or interstellar phases: 
\begin{itemize} 
    \item Any linearly polarized extragalactic (EG) source will provide a RM value, containing contributions from the Milky Way, the intergalactic medium, any intervening structures, and the environment in the source itself. The intergalactic contribution is relatively small, and for all but the highest latitudes, the contribution from the Milky Way dominates. With large numbers of EG sources one can construct an {\it RM Grid} of irregularly spaced RM measurements across the sky. \citet{VanEcketal2023} provided a consolidated table of over 55,000 EG source RMs. \citet{Hutschenreuteretal2022} removed EG source contributions and provided a best-estimate all-sky map of Galactic Faraday rotation (`the Faraday sky'). 
    \item Galactic pulsars also emit linearly polarized emission and therefore can be used as sources for an RM Grid. Pulsar RMs do not contain an intrinsic RM distribution, diminishing uncertainties in the Galactic RM. In addition, pulsar distances are often known (although not always accurately), allowing some distant-dependent information mostly concentrated near the Galactic plane. The comparison between the RM of pulsars and extragalactic sources can provide constraints on the magnetic field by segmenting the line of sight based on the distances of the pulsars \citep[e.g.][]{Han2017}. This method has been used to constrain Galactic magnetic field models as well as local structures \citep[e.g.][]{Xu2019MNRAS.486.4275X}. With SKA's sensitivity, the number of available pulsar RMs is expected to increase dramatically to tens of thousands \citep{Xue2017PASA...34...70X}.
    \item Faraday tomography of extended Galactic synchrotron emission provides a glimpse towards magnetic field structure in 3D space. Broadband spectro-polarimetric surveys of the whole sky show complex Faraday spectra, which allow disentangling of various Faraday rotation components along a line of sight ~\citep[e.g.][]{VanEcketal2017}.    
    \item The MC-BLOS technique of \citet{Tahanietal2018} can provide the line-of-sight component of magnetic field (\blos ) associated with molecular clouds by comparing EG RMs projected on or near a cloud versus slightly off the cloud, and performing chemical modeling to infer the thermal electron distribution in the cloud using column density maps. The selection of OFF points considers different criteria discussed in \cite{Tahanietal2025}.
    
\end{itemize}

\vspace{-5pt}
\subsubsection{Zeeman effect}
The Zeeman splitting of the atomic and molecular spectral lines in the presence of magnetic fields enables the measurement of the \blos . This technique provides the least model-dependent (and likely the most accurate) approach to determine the field strength. The emission or absorption from these split energy levels produces left- and right-hand circular polarization, the detection of which enables measurement of both the strength and direction of \blos . Zeeman observations of \HI\ and molecular tracers (such as OH, CN, CCS, and others) can probe magnetic fields across a wide range of densities and physical conditions in the interstellar medium \citep{Trolandetal1986, Crutcher1999APJ514, Crutcher2012, Sarmaetal2000, CrutcherKemball2019, Chingetal2022Nature}. The SKA will enable deep Zeeman observations of atomic and molecular tracers through its enhanced sensitivity and spectral resolution. For detailed technical specifications and science cases, we refer readers to two parallel chapters by \cite{Bourke01.2026.SKA} and \cite{Robishaw01.2026.SKA} in this Volume.

\subsection{Complementary/Ancillary commonly-used observational techniques}
\label{sec:other_methods}

The advent of SKA data is expected to enable the further development of the methods outlined in Section \ref{sec:ska-methods}, while at the same time allowing for new methods to be created. We anticipate that the combination of these methods with complementary probes of the magnetic field will provide fruitful ground for such new developments.

\subsubsection{Complementary radio-polarimetric surveys}
\label{sec:radio_surveys}

For extended synchrotron emission, single-antenna surveys can provide crucial information on large spatial scales, complementary to the high resolution of interferometer arrays (see \ref{subsubsec:spatialscales}). One key set of broadband, single-antenna extended emission polarization data is the Global Magneto-Ionic Medium Survey (GMIMS), covering 300--1800~MHz. GMIMS has two sub-bands completed in the northern hemisphere \citep{Wolleben2021, Ordog2026} with telescopes at the Dominion Radio Astrophysical Observatory, and two in the southern hemisphere \citep{Wolleben2019, Sun2025} with CSIRO's Murriyang Parkes telescope. Observations for a third southern hemisphere component, the POSSUM-EMU-GMIMS All Stokes UWL Survey (PEGASUS), have recently been completed with the Murriyang telescope. Currently existing and in progress interferometric datasets will also be useful for comparison with SKA data. These include the POlarization Sky Survey of the Universe's Magnetism \citep[POSSUM,][]{Gaensleretal2025}, with the SKA precursor, ASKAP, and the LOFAR Two-metre Sky Survey \citep[LoTSS,][]{shimwelletal2022} with the Low Frequency Array (LOFAR), an SKA pathfinder.

\subsubsection{Polarized thermal dust emission}
\label{section:dust}

Interstellar dust grains are typically non-spherical and spin around their minor axes. In the presence of magnetic fields, the axis of rotation can become statistically aligned with the local magnetic field direction
~\citep{purcell1975,Anderssonetal2015, HoangLazarian2016}, resulting in polarized thermal emission and extinction. 
The polarization orientation of dust emission observed at far-infrared and submillimeter wavelengths therefore traces the \bpos\  via a mass-weighted \citep{Seifriedetal2020} average along the line of sight. 

This technique has proven highly successful in mapping magnetic field morphology across diverse astrophysical environments, from the Galactic magnetic field structure studied by Planck observations \citep{Planck2016XLII} to magnetic fields in nearby galaxies~\citep{Borlaffetal2023, LopezRodriquezetal2023, Martin-Alvarezetal2024}, star-forming regions~\citep{Doietal2021Pinched, Tahanietal2023}, and protoplanetary envelopes and disks~\citep{Hwangetal2025, Guptaetal2022,  Sharmaetal2025}. Additionally, different far-infrared wavelengths probe different grain populations and physical conditions~\citep{Fanciulloetal2022}: shorter submillimeter wavelengths are sensitive to warmer dust closer to heating sources, while longer wavelengths trace cooler, more extended material, enabling magnetic field studies in different thermal environments within the same region. Additionally, dust polarization fractions can provide additional information about the inclination angle of the field with respect to the LOS ~\citep{Kingetal2018, Kingetal2019, Chenetal2019,Fisseletal2019, Sullivan2021,HuYue2023MNRAS.519.3736H,hoangtruong2024}.

\subsubsection{Starlight polarization}

Unpolarized starlight at optical and near-infrared wavelengths undergoes dichroic absorption as it passes through magnetically aligned dust grains, resulting in observed polarization aligned with \bpos . Additionally, since stellar distances are known through parallax measurements, particularly from Gaia~\citep{Lurietal2018}, this technique provides tomographic \bpos\ observations at different distances along the line of sight~\citep{Panopoulouetal2019, Doietal2021, Pelgrims2023A&A...670A.164P,Angaritaetal2025}. This distance information enables mapping of how the \bpos\ varies with depth, complementing the integrated measurements from dust emission. 
As with thermal dust emission, the polarization angle only provides the orientation of the \bpos , but not the direction, suffering from a 180$^{\circ}$ ambiguity. Similarly to thermal dust emission, stellar polarization can be used to infer the inclination angle of the magnetic field \citep{truonghoang2025}.

\subsection{Cross-Community Definitions}
\label{sec:definitions}
Across the astronomy and astrophysics community, some terminologies describing magnetic fields and interstellar structures can vary. For consistency, we highlight some of the terms and mention our descriptions of them in this section. We note that what may be referred to as ordered or turbulent field (also refered to as ordered random and isotropic random) in Galactic Magnetic field studies \citep{Jaffeetal2010, Jaffe2019} may be highly structured, coherent-looking fields in zoomed-in observations of specific regions, such as a star-forming region, where for example, gravity dominates and reshapes the field lines, or tangential fields around a bubble caused by their expansion.

For interstellar structures, we differentiate between \textit{\HII\ regions} (ionized hydrogen bubbles formed around massive stars, $\sim$ 1-10 pc), \textit{supernova remnants} ($\sim$ 10-100 pc shells of individual explosions), and \textit{superbubbles} ($\sim$ 100–500 pc often dust or hydrogen cavities caused by multiple supernovae). The term bubble may refer to any of the three above-mentioned shells. 
Throughout this chapter, the term \textit{filament} can refer to any elongated density structures, where ``\HI\ filaments’’ are mainly composed of neutral hydrogen in the diffuse ISM. \HI\ filaments can appear along the edges of bubble shells.  ``Molecular filaments’’ are dense, star-forming filamentary structures bright in CO observations. We use ``molecular clouds'' (which includes the term giant molecular clouds) for cold, molecular-hydrogen-dominated gas structures ($\sim$ 1–100 pc). 
In this chapter ``dense cores'' denotes sub-parsec regions where stars form with densities higher than $10^6$\,cm$^{-3}$. 

Finally, the diffuse medium includes interstellar structures with densities lower than densities of molecular regions $ \sim 500$\,cm$^{-3}$, and dense medium refers to structures with molecular line emissions and typical densities higher than $\sim 500$\,cm$^{-3}$. 

\section{SKA-Specific Observations and Technical Details}
Here we describe the properties of SKA AA4 for the application of Faraday tomography of extended synchrotron emission and RM grids in the context of previous surveys. Extended synchrotron emission may also be referred to as diffuse synchrotron emission in the literature. We adopt the term ``extended'' throughout this chapter to avoid confusion with the diffuse medium defined in Section~\ref{sec:definitions}. 

\subsection{Extended synchrotron emission }

\subsubsection{Faraday Depth Spectra with the SKA}
\begin{figure}
    \centering
    \includegraphics[scale = 0.37]{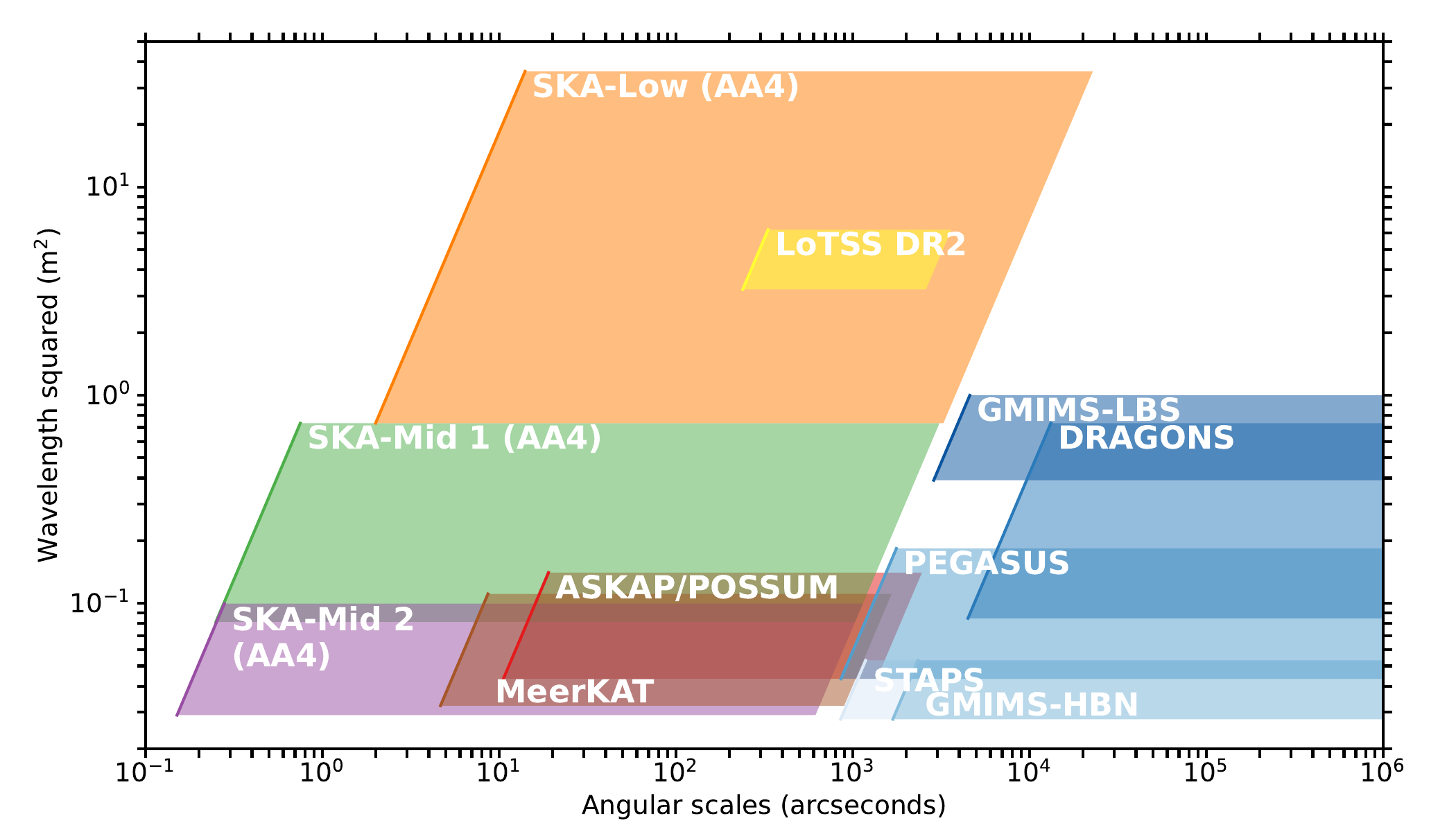}
    \caption{Angular resolution and wavelength squared coverage of existing and future interferometric and single-antenna surveys.}
    \label{fig:surveys_bw_res}
\end{figure}

The sensitivity to different Faraday rotation structures is characterized by three key parameters: the Faraday depth resolution, $\delta\phi$, the largest detectable Faraday depth, $\phi_{\text{max}}$, and the broadest feature in Faraday depth that is not depolarized, $\phi_{\text{max scale}}$. From \cite{BrentjensDeBruyn2005} these depend on the observing wavelength, $\lambda$, as
\begin{eqnarray}
\delta\phi\approx\frac{2\sqrt{3}}{\Delta\lambda^2},\\
|\phi_{\text{max}}|\approx\frac{\sqrt{3}}{\delta\lambda^2},\\
\phi_{\text{max scale}}\approx\frac{\pi}{\lambda_{\text{min}}^2},
\end{eqnarray}
where $\Delta\lambda^2$ and $\delta\lambda^2$ are the wavelength-squared coverage and sampling respectively, and $\lambda_{\text{min}}$ is the shortest observed wavelength. The ability to distinguish between multiple discrete Faraday rotating screens and broadened Faraday depth structures arising from mixed emission and rotation relies on observations having $\phi_{\text{max scale}}>\delta\phi$. This requires high frequencies in order to minimize $\lambda_{\text{min}}$, and a wide $\lambda^2$ coverage. The $\lambda^2$ ranges of SKA-Low and SKA-mid in comparison with SKA pathfinders and precursors is shown on the vertical axis of Figure~\ref{fig:surveys_bw_res}. With SKA-low (50-350~MHz) in AA4, $\delta\phi$ as narrow as 0.1~rad~m$^{-2}$ will be possible, a factor of 10 improvement in Faraday depth resolution over LoTSS. In terms of $\phi_{\text{max scale}}>\delta\phi$, SKA-Mid~2 in AA4 will have sensitivity to Faraday depth structures as wide as 110~rad~m$^{-2}$.

\begin{figure}
    \centering
    \includegraphics[width=\hsize]{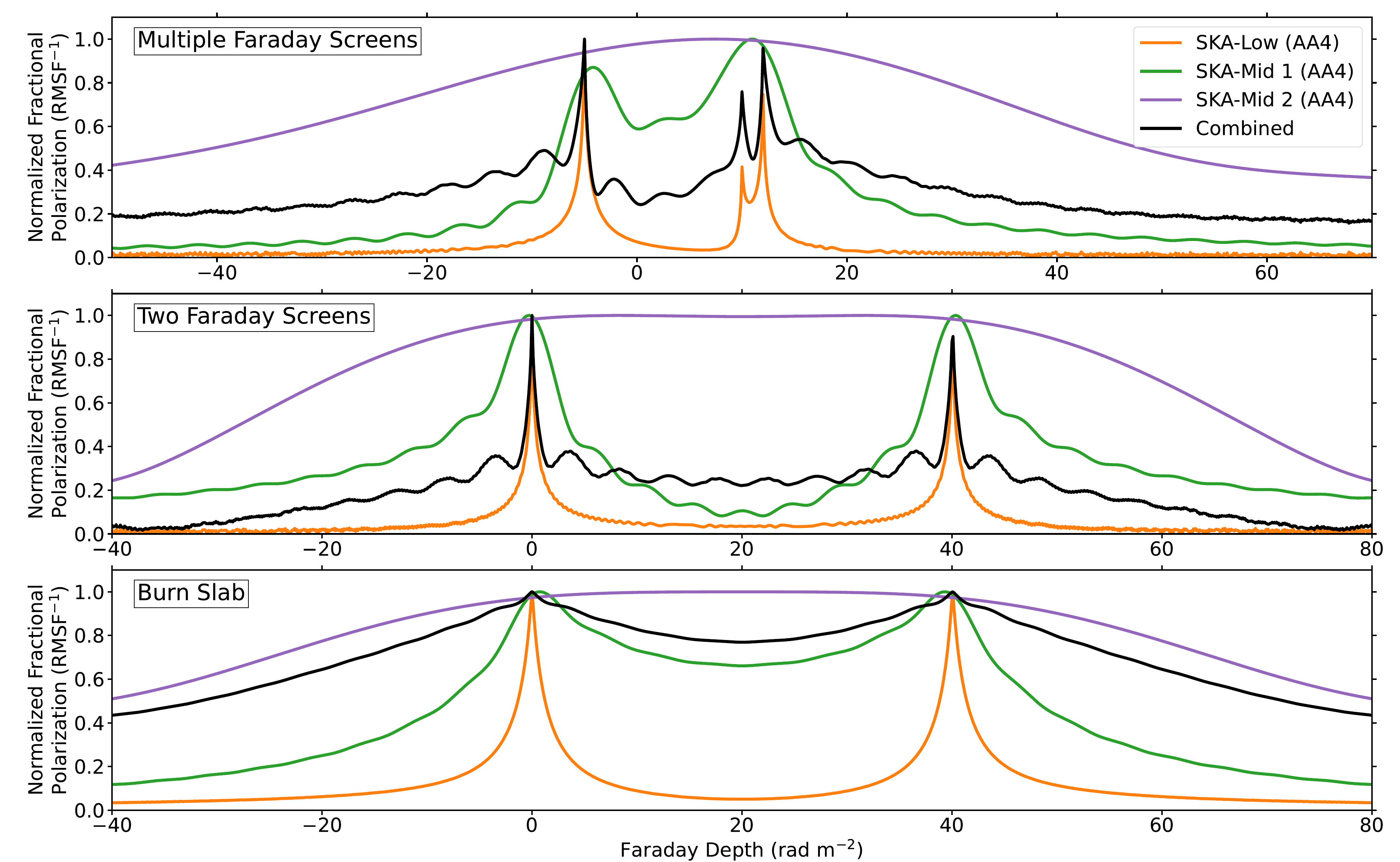}
    \caption{Examples of different Faraday rotation configurations with the separate SKA components, as well as combined. \textit{Top}: Multiple Faraday screens, resolved SKA-Low, partially resolved in SKA-Mid 1, and blended in SKA-Mid 2. \textit{Middle} and \textit{Bottom}: Comparing two well-separated Faraday screens to a Burn slab scenario with edges at the same Faraday depths. SKA-Mid 1 is able to distinguish between these scenarios.}
    \label{fig:example_spectra}
\end{figure}

Figure~\ref{fig:example_spectra} shows examples of ISM Faraday rotation scenarios and observed Faraday depth spectra with AA4 SKA-Low, Mid~1 (350-1050~MHz), Mid~2 (950-1760~MHz), and their combination. This highlights the ability of SKA-Low to resolve multiple, closely spaced Faraday depth features (top panel) and the role of SKA-Mid in distinguishing between Faraday screens and extended features similar to a Burn slab (middle and bottom panels). 
Given the differences in LOS depth probed by different frequency ranges (due to depth depolarization), examining Faraday depth spectra produced from specific frequency subsets will be informative. For example, combining only Mid-1 and Mid-2 data for comparison with Low-frequency observations can reveal how different frequency ranges probe different physical depths, especially when used in conjunction with other observables that help associate Faraday depth features with physical distances.

\subsubsection{Spatial Scales with the SKA}\label{subsubsec:spatialscales}
The horizontal axis of Figure~\ref{fig:surveys_bw_res} shows the range of angular scales of SKA and its precursors and pathfinders over their respective $\lambda^2$ ranges. For the interferometric data (SKA, ASKAP, MeerKAT and LoTSS DR2) the upper bound of this range is determined by the shortest available baselines and the lower bound is the resolution of the synthesized beam. At 140~MHz, SKA-Low will achieve $5\arcsec$ resolution, a factor of 60 improvement over LoTSS, making it ideally suited for studying extended synchrotron emission structures associated with \HI\ filaments \citep{VanEcketal2017, Braccoetal2020} and other diffuse ISM features. Faraday tomography requires spatially convolving all Stokes $Q$ and $U$ maps in the data cube to the angular resolution of the lowest frequency. At the lowest frequencies Mid-1 (Mid-2) will have $0.75\arcsec$ ($0.28\arcsec$) resolution, a factor of $\sim$12 (30) improvement over MeerKAT. 

Interferometric and single-antenna data are equally important and are complementary for studying extended synchrotron emission and Faraday rotation. For the synchrotron emission, single-antenna telescopes provide information on the largest spatial scales, providing the overall brightness for any given region, and allowing for the synchrotron emissivity to be calculated. Interferometric observations typically provide high-resolution information on the small-scale structures in an observed region, but do not observe the zero-level of the emission. For studying structures on a wide range of spatial scales in total and polarized emission, combined interferometric and single-antenna data are ideal. 
For extended emission Faraday rotation, analyzing combined interferometric and single-antenna data yields different information than analyzing either dataset alone~\citep{Ordog:2025}. When sufficient variability exists in the complex polarized emission to produce detectable interferometric signal, interferometric data can be sensitive to Faraday depth structures on spatial scales larger than the maximum nominal scale probed by the array~\citep[e.g.][]{Haverkorn:2003a, Ordog:2017}. In these cases, the extended emission Faraday rotation maps from interferometric data may correlate better with point source RM grids than with single-antenna maps~\citep{Ordog:2019, Erceg:2022}, as single-antenna observations are more strongly affected by beam depolarization.

For SKA AA4, GMIMS can provide complementary single-antenna data. The published and in-progress GMIMS data sets are shown in blue in Figure~\ref{fig:surveys_bw_res}. In particular, PEGASUS, for which observations are complete, is intended to provide large-scale information for POSSUM, but also overlaps in frequencies with SKA~Mid-1 and Mid-2, and could provide adequate large-scale information despite the small gap in spatial scale coverage. Similarly, the Southern Twenty Centimeter All-sky Polarization Survey \citep[STAPS;][]{Sun2025} will be able to provide large-scale polarization data for SKA~Mid-2. 

\subsection{The SKA RM Grid}

The SKA Cosmic Magnetism Working Group has proposed an observing plan in which an RM Grid plays a major role \citep{johnstonhollittetal2015,healdetal2020}. The RM Grid would be based on a primary polarizations survey using SKA1-MID Band 2, in the frequency range from 950 to 1760~MHz. They will result in a nominal RM precision $\Delta\mbox{RM}\lesssim 5$~rad~m$^{-2}$ for sources with a signal-to-noise ratio higher than 5, an expected $\gtrsim 100$ polarized sources per square degree, a factor~3 more than ASKAP \citep[see the parallel chapter by][]{Sun01.2026.SKA}, and a high angular resolution of about 2\arcsec.

\section{Diffuse ISM: Bubbles (Superbubbles,  Supernova Remnants, \HII\ regions) and Filaments}

SKA data will be invaluable for tracing magnetic fields in a variety of diffuse ISM structures, shaping our understanding of how these structures evolve and how they fit into the larger Galactic context. Massive stellar outflows, stellar UV radiation, and supernova explosions are among the dominant sources of energy injection into the ISM. These processes generate bubbles and compress the surrounding medium, which can trigger gravitational collapse and star formation~\citep{Deharvengetal2005}.  Magnetic field observations of these shells suggest that the fields are ``frozen-in'' to the medium~\citep{KothesBrown2009, Tahanietal2023}. The shells of bubbles often exhibit filamentary morphology~\citep{Heiles1979}.

\subsection{Supernova Remnants and Superbubbles}

A single supernova sweeps up the ISM, compressing the ambient magnetic field and causing it to wrap around the shell tangential to the surface of the bubble \citep[scales of $\sim$tens of pc;][]{KothesBrown2009, KimOstriker2015}. These field lines retain a memory of the initial (pre-explosion) field geometry while forming a morphology tangential to the bubble surface.
Inhomogeneities in the environment, such as density gradients, variations in initial field direction, and changes in magnetic field strength and direction, can lead to asymmetries in the supernova remnant shell \citep{Chenetal2017}. 
Multiple supernova explosions and/or outflows from OB associations~\citep[clusters of massive stars;][]{HigdonandLingenfelter2005, Bally2008, Collischonetal2021} within a localized region can create superbubbles (scales of $\gtrsim 100$~pc).  We suggest several techniques for reconstructing the 3D magnetic field geometry of these bubbles. 

\subsubsection{SNRs as Probes of the Galactic magnetic field}

\citet{2016A&A...587A.148W} showed that the morphology of SNRs are effective probes of their Galactic explosion environment, providing in-situ measurements of the Galactic magnetic field. The SNR morphology can reveal not only the orientation of the field, but also highlights gradients in the magnitude as well as bends and twists in its geometry. ASKAP observations of a relatively small region of $\sim35^\circ$ of the Galactic plane have added approximately 50 new radio SNRs to the known population \citep[including new identifications as well as new radio detections of candidates known at other wavelengths; ][]{2025ApJ...988...75B}. It is expected that by the end of the ASKAP-EMU survey, the population will grow by more than 400 candidates \citep{2025ApJ...988...75B}, doubling the known populations of SNR candidates in the Galaxy. SKA is expected to identify and confirm even more SNRs. The higher resolution will be useful for disentangling complex regions and identifying individual structures. This increased population will be particularly useful for developing the next generation large-scale 3D Galactic magnetic field models. 

\subsubsection{RM-Grid across a bubble}

RM-Grids are powerful tools to detect low-density ionized gas, and changes in the direction of the \blos . To date, the most commonly used RM-grid catalogue is that of \citet{Tayloretal2009}, which is derived from the 1.4 GHz NRAO VLA Sky Survey (NVSS, Condon et al. 1998) and the subsequent RMTable2023 \citep{VanEcketal2023}, has an RM-source density of only $\sim1$~per square degree. With such a low source density, we are restricted to analysis on large objects such as the Gum Nebula, which is a $36^\circ$ wide shell that is thought to be an old SNR \citep{2015ApJ...804...22P}. RM-grid densities that will be achievable by SKA will have 100 sources or more per square degree that will allow us to probe structures with arcmin scales. POSSUM has been used to probe a few extragalactic bubbles and supernova remnants \citep{2024MNRAS.534.2938J}. Stacking experiments could possibly be used to statistically measure the ionized gas density in the interior of the bubble. 

\subsubsection{Structure functions in the bubble walls}

\begin{figure}
    \centering
    \includegraphics[width=\hsize]{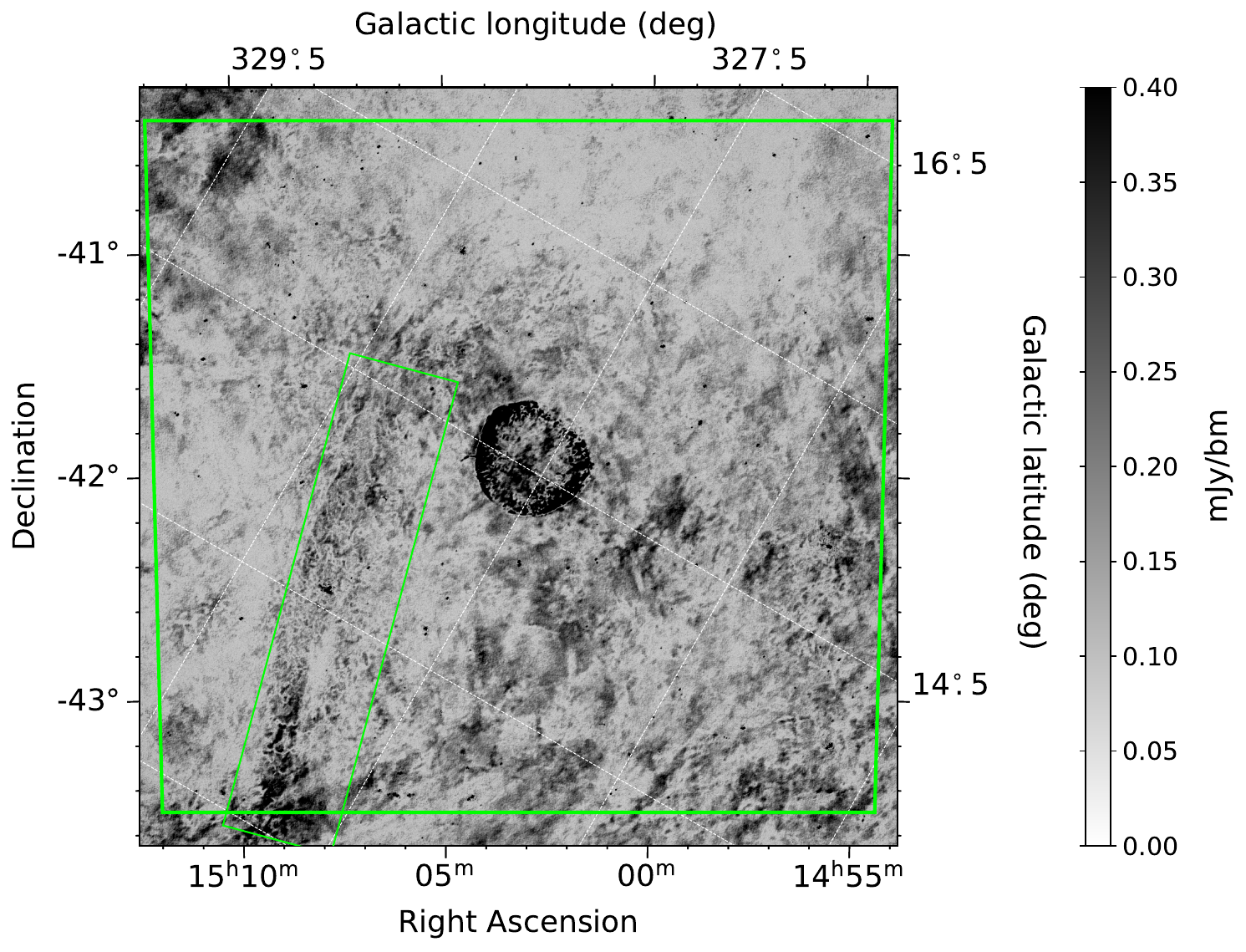}
    \caption{Peak polarized intensity map of the SN1006 field. The smaller green rectangle highlights a filamentary structure that has much finer scale structure when compared with the surrounding emission. (Reproduced with permission from \citet{Vanderwoude2025}).}
    \label{fig:sn1006}
\end{figure}

Large bubbles, such as superbubbles, with scales bigger than $\sim100$~pc, may exhibit different scales of structures in the front and back of the bubble wall, due to the differences in their distances. For example, \citet{Vanderwoude2025} used ASKAP observations with a $\sim18''$ beam to analyze a filamentary structure thought to be associated
with the Upper Centaurus Lupus (UCL) subregion of the Scorpius-Centaurus superbubble. The polarized intensity observations shown in Fig.~\ref{fig:sn1006} illustrate a clear difference in the scale of the polarization structure observed on and off the filament. If a structure with size 200~pc is located at a distance of 0.5~kpc, then variations on the order of 1~pc will appear as polarization structure with an angular size of $8.5'$ on the near side versus $6'$ on the far side. Such differences could plausibly be detectable with the arcsec resolution of SKA.

\subsubsection{Faraday tomography}

With the great improvements in Faraday depth resolution and sensitivity to structures, Faraday tomography may become a useful tool in distinguishing the front and back of a bubble. As the magnetic field is compressed and wraps around the shell of a bubble its direction changes along the line-of-sight for some viewing angles. In this way, the Faraday peaks of the bubble may be separated along the LOS, providing a unique probe of the 3D structure of the bubble \citep{2022MNRAS.513.3289I}. 
Correlating \HI\ emission of the bubble walls with diffuse polarized emission can help reveal the 3D structure. For this to work,  high spatial and velocity resolution are needed. 

\subsection{\HII\ regions}

Dust polarization observations of \cite{Tahanietal2023} using the James Clerk Maxwell Telescope have shown that \HII\ regions can push the magnetic fields and orient them tangential to the shell of these bubbles. They presented a formalism to use the field geometry and explore the feedback effects on magnetic fields. This formalism could potentially be extended to examine the energy budget in \HII\ envolopes and better examine 3D magnetic fields. A combination of these \bpos\ observations with Zeeman ~\citep{Sarmaetal2000}, and/or Faraday tomography and RM grids, could help reveal the 3D magnetic fields associated with the bubble surface of \HII\ regions. 

\subsection{\HI\ filaments}

The ISM exhibits filamentary morphology across a wide range of densities \citep{Hacar2023ASPC..534..153H}. The diffuse neutral ISM traced by \HI\ appears highly magnetized and filamentary, with filaments showing a statistically preferred orientation along the POS magnetic field as traced by starlight polarization and polarized dust emission \citep[e.g.][]{Clarketal2014}. At high Galactic latitude, \HI\ filaments are likely cold neutral medium (CNM) structures \citep{Clark2019ApJ...874..171C,Kalberla2025A&A...694L..11K}, the majority of which are within a few hundred parsecs from the Sun, usually associated with the wall of the Local Bubble \citep{Kalberla2016ApJ...821..117K}.

Polarized thermal dust emission and starlight polarization towards such \HI\ filaments can be used to obtain estimates of the POS magnetic field of these structures via the Davis-Chandrasekhar Fermi technique and its potentially more accurate variants \citep{Davis1951PhRv...81..890D,ChandrasekharFermi1953, Hildebrandetal2009, Houdeetal2009, SkalidisTassis2021}. In particular, \cite{Skalidis2019A&A...631L..11S} suggest that the Planck 353 GHz polarized emission at high latitude is mostly dominated by the Local Bubble.  
While this finding could allow for Local Bubble magnetic field estimates without correcting for background emission, \citet{Halaletal2024} presented results that could indicate otherwise, requiring multiwavelength observations. 
Such estimates can be supplemented by optical/NIR polarization of nearby stars. 

The LOS component of the magnetic field towards \HI\ filaments is most directly probed by observations of the Zeeman effect in circular polarization of the \HI\ line emission or absorption \citep[see parallel chapter by][]{Robishaw01.2026.SKA}. Magnetic field properties will of course vary as a function of scale, making 3D reconstruction of field vectors from the combination of dust polarization and the Zeeman effect difficult in general (e.g. due to small-scale tangling and reversals of the field). However, \HI\ filaments in the Local Bubble wall may offer an `easy' testbed: the magnetic field appears to be coherent enough from scales of few arcminutes up to a degree \citep{Nowotka2025arXiv250820065N}. 

SKA-Low data will also offer an alternative way of probing LOS magnetic fields in \HI\ filaments via the RM synthesis of extended low frequency radio emission. This technique has been widely used in LOFAR data revealing structures in Faraday depth space that appear to be correlated with tracers of the neutral medium \citep{Zaroubi2015MNRAS.454L..46Z,Kalberla2016A&A...595A..37K, Jelic2018A&A...615L...3J, Turic2021A&A...654A...5T, Bracco2020A&A...644L...3B}, often associated with the Local Bubble wall \citep{Erceg2024A&A...688A.200E}. \citet{Boulanger2024A&A...687A.102B} found that the local (within 200 pc) warm neutral medium (WNM) is the likely source of free electrons causing Faraday rotation observed in the LoTSS survey. 
To the extent that the magnetic field strength is constant with density in the diffuse neutral ISM \citep{Crutcher2012}, or at least has only a mild dependence with density \citep{Ponnada2022MNRAS.516.4417P,Seta2025MNRAS.539.1024S}, such low-frequency-based Faraday depths could be used to constrain the line of sight component of the magnetic field in nearby CNM structures.

The high density of extragalactic RMs could enable the estimation of the LOS magnetic field in distant, more massive \HI\ filaments \citep[e.g.][]{Syed2022A&A...657A...1S}, by extending the ON-OFF extragalactic RM technique (see Section \ref{sec:blos}) \citep{Tahanietal2018} to atomic gas. Any spatial gradient in the associated RMs can be compared to the filament axis and could be used to inform 3D models of the filament's magnetic field. Such measurements would be invaluable for comparing with GMF models \citep{Jaffe2019}, providing new constraints on the scale-dependence of the magnetic field properties at various galactocentric radii.

SKA data will allow us to explore the combination of velocity and Faraday depth spectra, probing discrete magnetized structures along the line of sight in a tomographic way  (most likely to appear at relatively low Galactic latitudes). This can be done, for instance, by searching for Faraday depth peaks that are morphologically similar on the plane of the sky with \HI\ features at different radial velocities using techniques such as machine vision algorithms \citep{Soler2019A&A...622A.166S}. If such identifications are robustly made, one could further exploit spatial gradients in the Faraday depth features to infer 3D properties of the \HI\ filament magnetic field (e.g. testing uniform versus kinked/arc-shaped field geometries). Informed by models of Galactic rotation, such estimates can be localized in 3D space, and further compared with complementary techniques of magnetic field estimation \citep[e.g.][]{Hu2023MNRAS.524.2379H}.

With \blos\ and \bpos\ estimates towards a filament, the inference of the 3D geometry of the magnetic field can be pursued. 
Different tracers of the magnetic field have a different dependence on the 3D geometry of the field, leading various authors to propose that combinations of multiple magnetic field tracers can be used to infer the 3D geometry \citep[e.g.][]{Goodman1994ApJ...424..208G,Reissletal2018}. 
However, complications arise due to the 180-degree directional ambiguity in \bpos\ \citep{Tahani2022} and the multi-phase, turbulent nature of the ISM. 
Improved methods of 3D inference of the magnetic field based on combinations of observables and statistical treatment of the turbulent magnetic field should be further developed \citep[e.g.][]{Tsouros2024A&A...690A.102T}, in order to fully exploit the wealth of magnetism data from the SKA.

\section{Dense ISM: Molecular clouds, filaments, and cores}

\subsection{Molecular clouds and filaments}

Magnetic fields influence the formation and evolution of molecular clouds, however, the details remain poorly understood. Magnetic fields and turbulence can significantly influence star formation rates and efficiencies~\citep{KrumholzFederrath2019}, and may stabilize clouds against gravitational fragmentation~\citep{Inoueetal2018, FiegePudritzI2000, FiegePudritzII2000}. Studies of the relative orientation between magnetic fields and density structures reveal close interplay between gas dynamics and field orientations~\citep{PlanckXXXV}. 

The last decade of magnetic field observations in molecular clouds have revealed a systematic trend: magnetic fields tend to align perpendicular to high-density structures ($>10^{21.7}$ cm$^{-2}$) and parallel to low-density structures~\citep{PlanckXXXV, Soler2019A&A...622A.166S}. At very high densities, another transition back to parallel alignment with dense filaments has been observed~\citep{Pillaietal2020}, potentially indicating gravitational dominance over magnetic fields (gravity pulling the field lines along dense filaments). The physical origin of the parallel-to-perpendicular transition remains debated~\citep{Crutcher2012, Seifriedetal2020, Pattleetal2023PP7}, with proposed explanations including converging flows~\citep{SolerHennebelle2017}, super-Alfv\'enic transitions~\citep{Chenetal2020}, and gravitational domination~\citep{Seifriedetal2020}. However, \citet{Pattleetal2023PP7} found no conclusive evidence linking the transition to specific Alfv\'en Mach numbers in their review.

Studies exploring field-density alignment examine only projected quantities on the plane of the sky. As demonstrated in Figure~\ref{fig:2DLimitationCartoon}, nearly perpendicular structures in three dimensions can appear parallel when projected, leading to incorrect physical interpretations. 
Additionally, true 3D magnetic field information is essential for accurately calculating physical quantities such as Alfv\'en Mach number, magnetic pressure, and mass-to-flux ratio, all of which require knowledge of the complete 3D field geometry.

In the following subsections, we describe techniques for determining line-of-sight and plane-of-sky magnetic field components in molecular clouds, and approaches for reconstructing complete 3D field vectors.

\subsubsection{Line-of-Sight Fields} 
\label{sec:blos}

\citet{Tahanietal2018} demonstrated that Faraday rotation measurements of unresolved background sources (pulsars or extragalactic sources) can be used to infer \blos\ associated with molecular clouds. The technique uses OFF points, which are sources whose projections fall near the cloud but sufficiently far from it in order to estimate the Galactic contribution (foreground and background to the cloud, i.e., non-cloud components) to the RM. The cloud's contribution is then isolated from ON points, which are sources with their projection either on the cloud or near enough to the cloud that their rotation measures are influenced by the cloud's magnetic field. Subtracting the Galactic RM (determined from OFF points) from the observed RM at ON points yields the cloud's RM contribution, revealing the direction of \blos . To determine field strength, the technique incorporates extinction or column density maps with chemical evolution models. In this approach, the cloud is divided into extinction layers along the line of sight, and electron densities are estimated at each layer as detailed in~\citet{Tahanietal2018, Tahanietal2025}. This approach accounts for the varying ionization fraction with depth into the molecular cloud.

\citet{Tahanietal2025} developed the MC-BLOS software package to implement this technique in an automated manner for upcoming RM observations from SKA and its pathfinder's surveys such as POSSUM~\citep{Gaensleretal2025} and SPICE-RACS~\citep{Thomsonetal2023SPICERACS}. The \blos\ results show excellent consistency with available Zeeman observations. 
In molecular regions, MC-BLOS results match very well in both strength and direction with available molecular Zeeman measurements. In cloud envelopes, MC-BLOS \blos\ directions match well with Zeeman, though the strength values may exceed the strengths estimated through atomic Zeeman measurements. 
However, the spatial correlations in the Orion A cloud envelope are striking: both the error-weighted atomic Zeeman and MC-BLOS-derived LOS field strengths on the eastern side of the cloud (where both techniques indicate fields pointing toward the observer) are twice the averaged error-weighted values on the western side (where both techniques indicate fields pointing away from the observer).

The SKA era will enable significant improvements to this technique through:
\begin{itemize}
\item \textbf{Better OFF point selection}: Current RM catalogs provide only $\sim 1$ source(s) per square degree, limiting OFF point selection~\citep{Tahanietal2025}. SKA's RM source density ($\sim 100$ sources per square degree for the RM Grid survey, potentially much higher for deep observations) will enable much more accurate determination of Galactic RM contributions. Pulsar RM observations will additionally aid Galactic RM modeling.
\item \textbf{Higher spatial resolution}: SKA's RM source density will produce much higher resolution \blos\ maps and extend the technique to many more molecular clouds, enabling refinement through examination of diverse regions.
\item \textbf{Extension to complex regions}: Combined with 3D dust maps and tomographic pulsar RM analysis, the technique can be applied to regions closer to the Galactic plane and more complex environments (with multiple clouds along the line of sight) than previously accessible.
\end{itemize}

\begin{figure}
    \centering
    \includegraphics[scale=0.27]{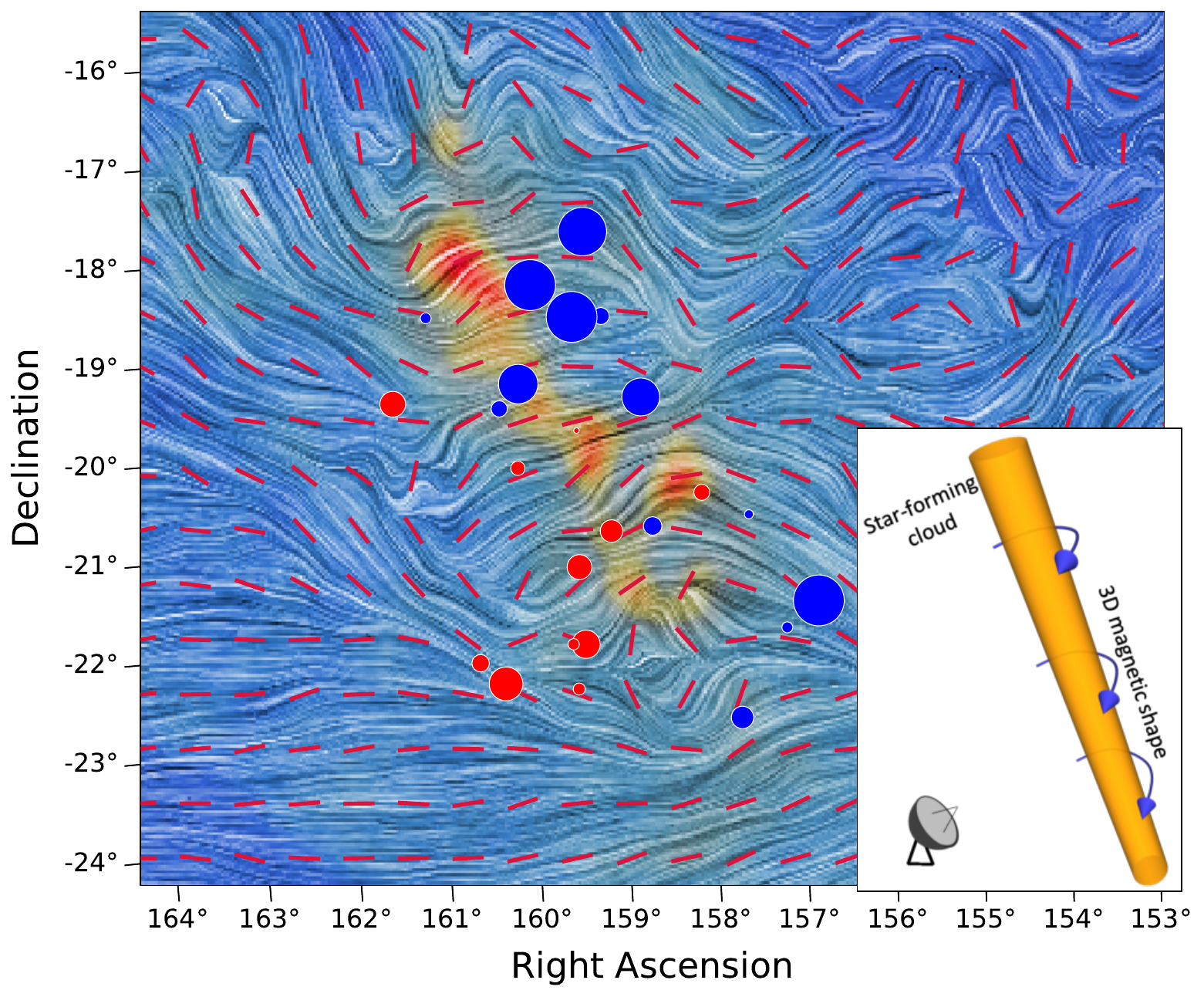}
    \includegraphics[scale=0.34]{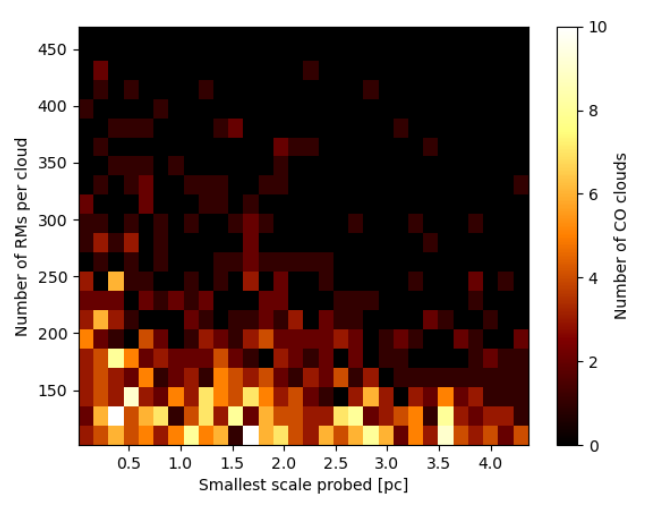}
    \caption{Left: Line-of-sight magnetic fields obtained by the MC-BLOS technique in the Perseus cloud and its reconstructed 3D field, bending around the cloud and forming a concave arc-shaped morphology \citep{Tahanietal2022P}. Right: Expectations for probing molecular clouds at low latitude with an RM grid: 2D distribution of the number of RMs per cloud versus the corresponding smallest POS scale that will be probed. Calculations based on an RM grid density of 100 sources/deg$^2$ and the \citet{Miville2017ApJ...834...57M} catalog of molecular clouds in the Galaxy (see text). }
    \label{fig:rms_per_cloud}
\end{figure}

With 100 RMs per square degree, we estimate the number of points per cloud will be hundreds for the majority of nearby molecular clouds. For the Perseus cloud, which covers $\sim 30$ square degrees, the SKA RM grid will provide thousands of measurements. So far, the MC-BLOS technique has been applied to high-latitude clouds with minimal confusion along the LOS. However, the high density of RMs provided by the SKA will allow extending the technique to lower latitudes. We can estimate the number of expected measurements per cloud, as well as the minimum lengthscale that will be probed on average, by making use of the most recent, complete catalog of molecular clouds in the Milky Way \citep{Miville2017ApJ...834...57M}. We select clouds that are within 5 kpc, covering at least 1 sq. deg of sky area, in the Southern sky ($\delta < 30^\circ$).  These are all clouds within $5^\circ$ of the galactic plane \citep[defined by the coverage of the original survey by][]{Dame2001ApJ...547..792D}. Assuming a random (Poisson) background source sky distribution, the mean separation between nearest neighbors is $\frac{1}{2\sqrt{100}} = 3'$. Figure \ref{fig:rms_per_cloud} (right) shows a 2D distribution of the number of RMs per cloud versus the smallest POS scale that can be probed per cloud with the assumed RM grid density. The expectation is that we will be able to probe LOS magnetic field fluctuations down to sub-parsec scales for a significant amount of clouds (given proper treatment of LOS-confusion effects). For further discussion on the capabilities of SKA in constraining small-scale magnetic fields in the Milky Way and nearby galaxies, we refer the reader to the chapter by \cite{Ma01.2026.SKA}.

In addition to RMs, Zeeman splitting measurements provide a powerful way of probing the \blos\ in molecular clouds through observations of molecular tracers such as OH and CN, as well as \HI\ to probe cloud envelopes. While precise g-factor determinations are essential \cite[see the parallel Zeeman chapter by][]{Bourke01.2026.SKA}, Zeeman measurements provide the least model-dependent (and likely the most accurate) technique for determining magnetic field strength. SKA observations can significantly expand available Zeeman measurements of molecular clouds through both absorption and emission detections. These observations are particularly valuable for tomographic studies of \blos\ when combined with MC-BLOS, as different tracers probe distinct density regimes: Zeeman measurements sample localized regions dominated by the density regime of each specific tracer, and MC-BLOS provides cloud-averaged \blos . Together with Faraday rotation measurements from extended emission and RM grids, multi-tracer Zeeman observations can enable tomographic mapping of \blos\ variations along the line of sight.

\subsubsection{3D vectors and LOS and POS integration}  
As described in Section~\ref{section:dust}, dust emission polarization and starlight polarization provide complementary measurements of \bpos\ in molecular clouds. 
Co-integration of  \blos\ and \bpos\ with models can help with reconstruction of 3D magnetic field vectors. \citet{Tahanietal2019} constructed models of the Orion A magnetic field morphology (that were consistent with \blos\ observations), generated synthetic observations from these models, and compared them with the \blos\ and \bpos\ observations. Using Monte Carlo simulations and chi-squared probability analyses, they concluded that an arc-shaped magnetic field, where the field bends around the cloud and connects to the larger-scale Galactic magnetic field, provides the most probable morphology and the best match to observations.

To resolve the 180-degree directional ambiguity in \bpos\ and determine complete 3D field vectors, \citet{Tahanietal2022O} and \cite{Tahanietal2022P} combined Galactic magnetic field models with \bpos\ and \blos\ observations. They used the Hammurabi code~\citep{Waelkensetal2009} with only the coherent component of the \citet{JanssonFarrar2012} Galactic field model, as the turbulent component is not relevant for determining the initial field geometry (before the evolution or formation of the cloud). Their reconstructions incorporated energy balance calculations including Alfv\'en Mach number and magnetic pressure. For the clouds they studied (Orion A and Perseus), the coherent components of various Galactic magnetic field models showed good agreement. Improved determination and modeling \citep[e.g.][]{UngerFarrar2024} of the large-scale Galactic magnetic field through SKA observations will enable extension of these complete field maps (six-dimensional: 3D vectors in 3D space) to many more molecular clouds.  

The combination of these techniques with stellar polarimetry and distance information from Gaia enables tomographic mapping of how \bpos\ varies with depth through molecular cloud complexes~\citep{Doietal2021} and how cloud fields connect to galactic magnetic fields. A number of surveys of optical/NIR linear polarimetry of stars are (or will become) available, targeting the inner Galaxy \citep{Clemensetal2020}, higher latitudes \citep[PASIPHAE,][]{tassisetal2018}, lower latitudes including the Galactic plane \citep[SOUTH POL,][]{magalhaesetal2018}, and targeted regions \citep[DragonflyPol;][Tahani et al., in prep]{Tahanietal2026}.

In addition to advancing our understanding of magnetic field roles in star formation, 3D magnetic field vectors can reveal the formation history and evolutionary pathways of molecular clouds. 
The reconstructed 3D magnetic fields of the Orion A and Perseus molecular clouds enabled step-by-step formation scenarios for these clouds, revealing influence from nearby bubbles consistent with the formation models of \citet{Inutsukaetal2015}.
For example, the 3D field vectors of the Perseus molecular cloud showed the influence of a previously unidentified interstellar structure on the cloud's formation and evolution, a structure invisible in total emission observations. This finding was subsequently confirmed through independent kinematic studies~\citep{Kounkeletal2022}.

\subsection{Dense Core Studies with SKA} 
\label{sec:cores}

\subsubsection{Zeeman Effect Observations of Dense Cores with SKA-Mid} 

The reconstruction of 3D magnetic field vectors threading molecular dense cores requires complementary measurements of both line-of-sight and plane-of-sky field components. As discussed in Section \ref{sec:other_methods}, polarized dust emission and starlight polarization trace \bpos\ 
through mass-weighted and tomographic observations, respectively, while Zeeman splitting provides direct measurements of \blos . 
For dense molecular cores, the integration of these techniques helps construct a whole picture of the cores from a magnetic field point of view, namely Zeeman observations of \HI , OH, and CCS transitions measure \blos\ through the core material itself, while Faraday rotation of background continuum sources and foreground/background pulsars provides complementary \blos\ 
constraints from the medium surrounding cores.

SKA-Mid will significantly improve both measurement techniques through unprecedented sensitivity, achieving an order of magnitude improvement over current interferometers while providing sub-arcsecond angular resolution. This will enable detection of magnetic fields down to (1--2)~\(\mu\)G in OH transitions and potentially sub-\(\mu\)G levels in \HI, under favorable observing conditions. This transformational capability addresses critical gaps in 3D field reconstruction: current facilities detect Zeeman signatures in only 25--30\% of targeted molecular cores with integration times of 10--50~hr, while SKA-Mid is expected to detect comparable fields in a few~hrs with detection rates exceeding $\sim$ 50\%.

\subsubsection{Sensitivity and spectral resolution for OH transitions}
All four ground-state OH transitions relevant for Zeeman effect studies (satellite lines at 1612 MHz and 1720 MHz, and main lines at 1665 MHz and 1667 MHz) lie within SKA-Mid Band 2 (950–1760 MHz), which provides up to 810 MHz of simultaneous bandwidth. These transitions can be observed simultaneously. 

\subsubsection{Specifications for CCS molecular line observations}

CCS presents a promising but challenging tracer for high-density regions. 
The key CCS transition for Zeeman measurements, the \( J_N = 1_0 \rightarrow 0_1 \) line at approximately 11.119~GHz \citep{shinnaga2000}, falls within SKA-Mid Band~5b coverage (8.3--15.4~GHz). At 12.5~GHz, representative of this band, SKA-Mid is expected to achieve an array System equivalent flux density (SEFD) of 2.77~Jy and \( A_{\mathrm{eff}}/T_{\mathrm{sys}} \) of 998~m\(^2\)/K, yielding a spectral line sensitivity of 85~\(\mu\)Jy/beam for one-hour integrations with the 210~Hz zoom channel resolution available across all bands.

CCS traces densities of \(10^4\)–\(10^5~\mathrm{cm}^{-3}\) in cold, dense pre-stellar cores \citep[e.g.][]{suzuki1992}. 
The Zeeman coefficient for CCS near 11~GHz is approximately 0.8~Hz/$\mu$G, making detectable field strengths around 200~$\mu$G feasible for strong lines. However, CCS observations face the following challenges: higher frequencies require better atmospheric conditions; the smaller beam size (down to $0.05''$) at 11~GHz samples small physical scales, potentially missing extended emission; and CCS emission is typically weak and spatially complex. SKA Mid's enhanced sensitivity may have a great potential 
in establishing CCS as a routine tracer of magnetic fields at higher densities.

\subsubsection{\HI\ 21-cm line observations: Emission and absorption as complementary dense core tracers}
\subsection*{Zeeman Observations in \HI\ Emission and Absorption}

\textbf{\HI\ emission Zeeman observations} probe magnetic fields in the cold neutral medium (CNM) surrounding molecular clouds, setting key boundary conditions for understanding magnetic field evolution during cloud formation and collapse. The Zeeman splitting of the \HI\ 21-cm line is 2.8~Hz\,$\mu$G$^{-1}$ (0.59~km\,s$^{-1}$\,mG$^{-1}$), comparable to OH splitting and resolvable with SKA-Mid's finest 210~Hz (0.044~km\,s$^{-1}$) spectral resolution. At 1.4~GHz, SKA-Mid achieves $\sim$140~$\mu$Jy\,beam$^{-1}$ line sensitivity in one hour, enabling detection of $\sim$0.5--1~$\mu$G fields in \HI\ emission with typical brightness temperatures of 20--100~K. Mapping diffuse \HI\ requires moderate angular resolution (10--60\arcsec) to avoid spatial filtering, favoring use of SKA's inner array or combination with MeerKAT, which offers sensitivity to large-scale emission while retaining interferometric suppression of systematics.

\textbf{\HI\ absorption observations} access different environments. 
This technique 
enables independent estimates of gas density, temperature, and magnetic field strength. Absorption is particularly powerful in dense core studies, detecting cold atomic gas at 10--30~K that is indistinguishable from warmer \HI\ in emission.

\textbf{\HI\ Narrow Self-Absorption (HINSA)} is a specialized form of absorption spectroscopy that isolates cold atomic hydrogen mixed with molecular gas \citep{Li+Goldsmith2003,Goldsmith+Li2005}. HINSA features are narrow ($\Delta v \lesssim 1$~km\,s$^{-1}$) absorption dips superimposed on broader \HI\ emission, produced by \HI\ at $T_{\rm spin} \sim 10$--40~K cooled via collisions with H$_2$. Unlike classical \HI\ self-absorption (HISA), which traces warm-to-cold transitions in atomic gas, HINSA identifies residual \HI\ within well-shielded regions dominated by H$_2$, with [\HI]/[H$_2$] $\sim 10^{-4}$--$10^{-3}$. Although volumetrically minor, this cold \HI\ component is an excellent magnetic probe because: (1) HINSA lines are 5--10$\times$ stronger than typical molecular tracers; (2) Zeeman splitting is as strong as in normal \HI\ (2.8~Hz\,$\mu$G$^{-1}$); and (3) \HI\ abundance is stable across diverse environments, avoiding depletion effects seen in molecules.

A HINSA Zeeman detection by \citet{Ching2022Natur.601...49C} with FAST toward L1544 in Taurus measured a field strength of $+3.8 \pm 0.3$~$\mu$G. This was part of a multi-tracer analysis combining quasar \HI\ absorption at 7.1~pc, \HI\ emission at 0.72~pc, OH emission at 0.24~pc, and HINSA at 0.15~pc from the core center. The data revealed a coherent magnetic field structure over four orders of magnitude in density—from CNM ($n \sim 30$~cm$^{-3}$) to molecular envelopes ($n \sim 10^4$~cm$^{-3}$) and dense cores ($n \sim 10^5$~cm$^{-3}$), with a consistent $\sim$4~$\mu$G field strength. 

\subsection*{SKA-Mid's Capabilities}

SKA-Mid will transform HINSA from a niche technique into a survey tool for magnetic fields in dense cores. With 2~$\mu$Jy\,beam$^{-1}$ continuum sensitivity at 1.4~GHz, SKA-Mid can detect HINSA against background sources with flux densities down to 10--50~mJy, three orders of magnitude fainter than required for single-dish detections. This extends HINSA Zeeman studies beyond nearby clouds ($d \ll 200$~pc) to targets across the Galactic plane ($d \sim 1$--2~kpc). The interferometric architecture offers key advantages: systematics suppression via cross-correlation, sub-arcsecond resolution to isolate compact continuum sources, and a large field of view enabling simultaneous emission and absorption observations across multiple sightlines.

\subsection*{Toward 3D Magnetic Field Mapping with SKA-Mid}

Combining SKA-Mid Zeeman observations in \HI\ (emission, absorption, HINSA), OH, and CCS enables comprehensive 3D magnetic field reconstruction across the full density range of star-forming regions. \HI\ traces $\sim$0.1--1~$\mu$G fields in diffuse CNM; OH detects 1--10~$\mu$G fields in molecular envelopes; HINSA probes 1--5~$\mu$G fields in atomic-molecular interfaces; CCS accesses 0.1--1~mG fields in dense prestellar cores. No other facility offers such a diagnostic suite with matched sub-arcsecond spatial resolution. The resulting datasets will allow systematic study of how field strength, geometry, and dynamical influence evolve from diffuse atomic gas through molecular envelopes to collapsing cores.

\subsubsection{Resolving core-envelope three-dimensional field structure}

Resolving the core-envelope 3D magnetic field structure is critical for understanding magnetic braking and angular momentum transport. Theoretical models predict that magnetic field lines become increasingly pinched as material contracts, potentially launching magnetically driven outflows via magneto-centrifugal processes or forming magnetically supported pseudo-disks where field lines resist further compression. Current observations largely measure single-pointing field strengths, inferring morphology from polarization angles \citep[][]{Kandorietal2017, Clemensetal2016, Kandorietal2020_3Dc} but lacking direct 3D vector measurements revealing field curvature.

SKA-Mid’s sub-arcsecond resolution can potentially enable mapping of \( B_{\rm LOS} \) from 1000~AU envelope scales to 100~AU inner core scales. When combined with dust polarization data, 
this could enable reconstruction of 3D magnetic fields. The ratio \( B_{\rm LOS}/B_{\rm POS} \) is expected to vary systematically with radius in flux-freezing scenarios, producing an hourglass morphology with field lines transitioning from poloidal in envelopes to toroidal in the flattened core. In contrast, turbulent field models predict randomized \( B_{\rm LOS}/B_{\rm POS} \) ratios at all radii. Furthermore, with resolved observations of the Zeeman effect towards cores, we will be able to detect gradients in the total field strength.  
This will enable detailed comparisons with theoretical predictions of the radial variation of the mass-to-flux ratio due to ambipolar diffusion and other non-ideal magnetohydrodynamics effects.

\subsubsection{Synchrotron emission from protostellar magnetospheres as complementary dense core diagnostics}

Beyond molecular line Zeeman observations, SKA-Mid's sensitive continuum capabilities enable the detection of non-thermal radio emission from embedded protostars, providing independent constraints on magnetic field strengths in the innermost regions of collapsing cores. Theoretical models and X-ray observations demonstrate that young protostars exhibit intense magnetic activity, with reconnection events and magnetospheric processes accelerating particles to relativistic energies \citep{Padovani2018A&A...620L...4P}. These energetic particles, gyrating in the protostellar magnetosphere's kilogauss-strength fields, produce synchrotron radiation detectable at radio wavelengths \citep{Bracco2025A&A...694A.148B}. Recent predictions suggest that Class~0 and Class~I protostars embedded in dense cores should produce synchrotron emission at flux-density levels of tens to hundreds of $\mu$Jy at GHz frequencies, with spectral indices $\alpha \sim -0.5$ to $-1.0$ characteristic of non-thermal emission and clearly distinguishable from dust thermal emission ($\alpha \sim +2$ to $+3$).

LOFAR observations targeting the \citet{Pezzuto2021A&A...645A..55P} catalog of high-density molecular cloud cores identified from \textit{Herschel} and \textit{Planck} surveys yielded non-detections of protostellar synchrotron emission, with $3\sigma$ upper limits of several hundred microjanskys at 150~MHz. However, these non-detections do not exclude the theoretical predictions, since synchrotron emission exhibits steep negative spectral indices and LOFAR's low frequencies probe the optically thick regime, where free-free absorption from ionized gas surrounding protostars can suppress the signal. SKA-Mid observations at 1--15~GHz access the optically thin synchrotron regime while providing order-of-magnitude sensitivity improvements over LOFAR at comparable frequencies. With continuum sensitivities of 2~$\mu$Jy\,beam$^{-1}$ at 1.4~GHz and 5~$\mu$Jy\,beam$^{-1}$ at 12~GHz in one-hour integrations, SKA-Mid can detect the predicted synchrotron emission from embedded protostars, enabling statistical studies correlating magnetic-activity signatures with core properties (mass, density, and magnetic field strength inferred from Zeeman measurements) and evolutionary state. The combination of Zeeman-derived large-scale field measurements and synchrotron-derived magnetospheric diagnostics will potentially provide unprecedented dual constraints on magnetic field structures from $\sim1000$~AU envelope scales down to $\sim1$~AU magnetospheric scales, testing theoretical predictions for magnetic flux conservation and field amplification during collapse. For an overview of observations of in-situ synchrotron emission from cores, we refer the reader to the parallel chapter by \cite{Bracco01.2026.SKA}.

\section{Summary}
\vspace{-6pt}
Mapping the 3D interstellar magnetic fields has far-reaching implications, from understanding galaxy evolution to illuminating the processes of cloud, star, and planet formation, cosmic-ray propagation, and beyond. In star-forming environments in particular, reconstructing 3D field vectors in 3D space enables detailed studies of energy budgets and a clearer understanding of the relative roles of gravity, turbulence, and magnetic fields in regulating star-formation rates and efficiencies. 3D field reconstructions allows us to trace how gas and dust flow to feed dense regions, how filaments fragment, and how magnetic geometry evolves from cloud to core. For example, compiling 3D fields of a large number of clouds, including isolated clouds such as Musca and bubble-impacted ones such as Perseus, enables exploration of how fields and clouds co-evolve to form stars in different environments and whether these two environments yield different star-formation rates and efficiencies ~\citep{Tahanietal2025PRIMA}.

Utilizing complementary techniques across the electromagnetic spectrum makes 3D field vector reconstruction possible. Polarization observations from radio to optical probe different components of the field and offer insights on different interstellar regions. However, major pieces of the puzzle have so far been missing. The SKA can provide those missing pieces. Its broadband, high-sensitivity spectro-polarimetric capabilities and dense rotation-measure grids will significantly improve measurements of the line-of-sight magnetic fields and the tomographic view of this field component along the line of sight. Additionally, SKA can enable improved plane-of-sky fields in synchrotron-dominated regions.

Combining upcoming data across the electromagnetic spectrum with SKA's radio measurements, we will be able to reconstruct complete 3D magnetic field vectors and localize them in space. Sub-millimeter and far-infrared facilities, such as CCAT-prime/FYST~\citep{CCATPrime2021} and potentially PRIMA~\citep{Tahanietal2025PRIMA} will map polarized dust emission and hence trace plane-of-sky field morphology in dusty regions. Optical and near-infrared polarization surveys such as DragonflyPol\footnote{\url{https://dragonflypol.github.io/DragonflyPol/}}~\citep[][Tahani et al., in prep]{Tahanietal2026}, PASIPHAE\footnote{\url{https://pasiphae.science}}, GPIPS \citep{Clemensetal2020}, and IPS \citep{Angaritaetal2025} provide tomographic views of plane-of-sky fields. These observations can provide field inclinations and can complement SKA observations.

The synergy of these facilities enables magnetic field studies over a vast variety of densities and temperatures. SKA's sensitivity and frequency coverage will open the door to mapping Faraday depth structures associated with bubbles and filaments, potentially detecting front- and back-shell signatures, which can be directly compared with dust-polarization or synchrotron-derived morphologies. SKA observations can bridge the gap between the Galactic magnetic field and the scales of individual molecular clouds and substructures within these clouds, in a co-investigation effort with other observations across the spectrum. Ultimately, these efforts will converge toward a comprehensive 6D view (3D field vectors mapped in 3D space) across different environments. This view can clarify how magnetic fields influence cloud formation and fragmentation, how feedback from stars and bubbles reshapes the surrounding interstellar medium, and how the Galactic magnetic field connects to star-forming clouds.

\section{Acknowledgements} \label{sec:ack}
G. V. P. acknowledges support from the Swedish Research Council (VR) under grant number 2023-04038 and the Knut and Alice Wallenberg Foundation Fellowship program under grant number 2023.0080.  
Claude.ai was used for editorial purposes. AI-assisted translation from Japanese to English was used for parts of Section~\ref{sec:cores}.

\bibliographystyle{abbrvnat-maxbibnames4}
\setlength{\bibsep}{0pt plus 0.2ex} 
\bibliography{ref} 

\end{document}